\documentclass[12pt,preprint]{aastex}

\slugcomment{Submitted to the Astronomical Journal}

\shorttitle{USNO-B}
\shortauthors{Monet et al.}

\begin{document}


\title{The USNO-B Catalog}


\author{
David G. Monet,
Stephen E. Levine,
Blaise Canzian,
Harold D. Ables\altaffilmark{1},
Alan R. Bird\altaffilmark{1},
Conard C. Dahn,
Harry H. Guetter\altaffilmark{1},
Hugh C. Harris,
Arne A. Henden\altaffilmark{2},
Sandy K. Leggett\altaffilmark{3},
Harold F. Levison\altaffilmark{4},
Christian B. Luginbuhl,
Joan Martini,
Alice K. B. Monet,
Jeffrey A. Munn,
Jeffrey R. Pier,
Albert R. Rhodes,
Betty Riepe,
Stephen Sell,
Ronald C. Stone,
Frederick J. Vrba,
Richard L. Walker\altaffilmark{1},
Gart Westerhout\altaffilmark{1}
}
\affil{U.S. Naval Observatory, Flagstaff, AZ 86002}

\author{
Robert J. Brucato,
I. Neill Reid\altaffilmark{5}
}
\affil{Palomar Observatory, Pasadena, CA 91125}

\author{
William Schoening\altaffilmark{1}
}
\affil{National Optical Astronomy Observatory, Tucson, AZ 85726}

\and

\author{
M. Hartley,
M. A. Read,
S. B. Tritton
}
\affil{Royal Observatory Edinburgh, Scotland, EH9 3HJ, UK}


\altaffiltext{1}{Retired}
\altaffiltext{2}{Universities Space Research Association}
\altaffiltext{3}{Joint Astronomy Center, Hilo, HI 96720}
\altaffiltext{4}{Southwest Research Institute, Boulder, CO 80302}
\altaffiltext{5}{Space Telescope Science Institute, Baltimore, MD 21218}


\begin{abstract}
USNO-B is an all-sky catalog that presents positions, proper motions,
magnitudes in various optical passbands, and star/galaxy estimators for
1,042,618,261 objects derived from 3,643,201,733 separate observations.
The data were obtained from scans of 7,435 Schmidt plates taken for the
various sky surveys during the last 50 years.  USNO-B1.0 is
believed to provide all-sky coverage, completeness down to $V = 21$,
0.2 arcsecond astrometric accuracy at J2000, 0.3 magnitude photometric
accuracy in up to five colors, and 85\% accuracy for distinguishing
stars from non-stellar objects.
A brief discussion of various
issues is given here, but the actual data are available from
{\em http://www.nofs.navy.mil} and other sites.
\end{abstract}




\section{Introduction}

The USNO-B catalog, currently released in version 1.0, is compiled from
the digitization of various photographic sky survey plates by the Precision
Measuring Machine (PMM) located at the U. S. Naval Observatory Flagstaff
Station (USNOFS).  The PMM produces both a pixel archive and a list of
detections for each plate in real time.
The plate material includes five complete coverages
of the Northern sky and four of the Southern sky, and contains a mixture
of colors and epochs.  The catalog was compiled from the merged lists
of detections, and presents position, proper motion, magnitudes in various
colors, star-galaxy classification, and various uncertainty estimators
for 1,042,618,261 distinct objects.  A discussion of the technical
details is presented in the documentation associated with the catalog,
and like the catalog is not being submitted for publication in a journal.
The digital distribution can be found at {\em http://www.nofs.navy.mil} and
other sites.

USNO-B is the next step in the sequence of catalogs that started with
UJ1.0 (Monet et al. 1994),
USNO-A1.0 (Monet et al. 1996) and USNO-A2.0 (Monet et al. 1998).  In
simple terms, USNO-A was a 2-color, 1-epoch catalog
while USNO-B is a 3-color, 2-epoch catalog.  This statement, like most
generalities about these catalogs, is incorrect.  The southern
surveys that were part of USNO-A were taken at 2 epochs,
and there is no first epoch blue survey south of $\delta = -33\arcdeg$
to include in USNO-B.  For both catalogs, an object must be detected on two
different surveys to be included in the catalog since isolated, single
survey detections are unreliable.  For USNO-A, this meant that the object
must have detectable red and blue fluxes, and this led to the exclusion
of many faint objects with non-neutral colors.  Also, the epoch difference
in the southern surveys meant that objects with larger proper motions
tended to be excluded.  USNO-B attempts to fix both of these problems.
An object detected on the same color survey at two epochs will be included
in USNO-B as will objects which have significant proper motions, although
it is still the case that objects with large motions and extreme colors
may be omitted.  The selection algorithm requires that spatially
coincident detections must be made on any two of the surveys
for an object to be classified as real and be included in the catalog.

\section{Plate Material}

Even in these modern times, one cannot help but be impressed by the effort
and dedication put into the photographic sky surveys.  A useful overview
was presented by Morgan et al. (1992), and Table~\ref{tbl-1} presents
the general properties of the surveys incorporated into the USNO-B catalog.
The following details are of general interest, but the rest are
relegated to the documentation associated with the digital version of
the catalog.

\begin{quote}
In perhaps 20 cases, the original plate (mostly in POSS-I) has
disappeared and cannot be found.  In these cases, the glass copy was scanned.
\end{quote}

\begin{quote}
It appears to take about 15 years to complete a photographic
survey.  The epoch for a particular field can be quite different from
the mean epoch, and this difference can be quite important in the
measurement of stellar proper motions.
\end{quote}

\begin{quote}
The total number of plates included in USNO-B1.0 is 7,435.  The
missing fields are numbers 1 and 646 in the SERC-I survey.  These were not
available in time for this release, but will be included in later releases.
\end{quote}

\begin{quote}
In our notation, there are 937 POSS-I fields.  The centers for
fields 1 and 2 and for fields 723 and 724 are the same.  Both fields 
1 and 2 are included since they were taken at very different hour angles
(they are at the North celestial pole), but field 723 was omitted since
724 was published in the atlas.  Because of these degeneracies,
different numbering schemes have been used over the years, and
some confusion persists.
\end{quote}

\begin{quote}
The plates involved in USNO-B are about 60\% of the Schmidt plates scanned
by the PMM.  Data from the other plates (rejected, Luyten's, etc.) will be
included in future USNO catalogs.
\end{quote}

\section{Scanning Procedure}

PMM differs from most other astronomical plate scanners because it
uses a 2-dimensional charge coupled device (CCD) camera as its sensor.
A Schmidt plate is digitized in 588 separate exposures, and each
produces a 1312 by 1032 image digitized with 8-bit resolution.  The CCDs
have 6.8 by 6.8 micron pixels, a 100\% fill factor, and are read at 10
million pixels per second.  A complete exposure cycle takes about 0.5
seconds including the shutter operation and integration times.   A lens
provides 2:1 imaging of the CCD on the focal plane.  Using a nominal
plate scale of 67 arcseconds per millimeter, the pixels are about
0.9 arcseconds in size, the field of view is about 20 by 15 arcminutes,
and spacing between exposures is about 18 by 13 arcminutes.  PMM has two
CCD camera systems, but mechanical limitations restrict parallel operations
to about 60\% of each plate.

PMM is also different from other scanners because all image processing
is done in real time.  Each CCD is connected to a computer by an optical
fiber and a DMA interface, and each frame is fully archived and processed
before the next is taken.  The processing is quite complicated, a
detailed description (including source code) is included in the USNO-B
documentation, but the following gives the major steps.  After the data arrive
in physical memory, the process forks and one path writes the raw data
to magnetic tape and then exits.  The processing path starts with bias
removal, flat field correction, and an object finder.  Each object
is processed in steps of increasing precision, but the processing of each
object is independent of all others.  This independence enables the usage
of parallel processing by the four CPUs in each computer, and was a
necessary simplification given the computers available in the early
1990s.  Table~\ref{tbl-2} lists all of the parameters computed for each
object in every frame.  This rich parameter set obviates the need to
make a ``star'' or ``galaxy'' decision in real time.
Once the entire plate has been scanned, the list of object parameters
is flushed first to disk, and then (offline) to CD-ROM.  All
data in the USNO-B catalog come from these files, and the lookback
index given for each detection in the catalog allows all
quantities to be recovered from the database.

Perhaps the only portion of the algorithm worthy of comment here is the
one used to compute the image position.  The image data are 8-bits
in transmission, and by measuring photographic negatives, the sky
is bright and the images are faint.  Since most of the survey plates
are lacking density calibration spots or wedges, and because the scattered
light in the images is small (perhaps 0.5\%) but significant, the choice
was made to omit a density-to-intensity calibration, and to deal
with the images in transmitted light.  All of these effects increase
the already noisy nature of the photographic emulsion, and tests showed that
a least squares estimator was necessary.  In the faint limit, stars look
something like a Gaussian, but they have flat cores and wide wings
when saturated.  The choice was made to use the function
\begin{equation}
T(x,y) = B + {A \over {e^{\alpha (r^2 - r_0^2)} + 1}}
\end{equation}
where $r^2 = (x-x_0)^2 + (y-y_0)^2$.  The six free parameters are
$A$ (central amplitude), $B$ (background), $x_0$, $y_0$ (position),
$r_0^2$ (saturation radius), and $\alpha$ (extent of image wings).
Extensive testing indicated
that this symmetric version provided a more stable estimate of the image
position than the somewhat better fitting function that includes a
seventh free parameter for asymmetry.
Traditional non-linear least squares based on
Marquardt's algorithm (Bevington 1969) was used.  Iteration was
terminated after three steps, regardless of convergence: testing
showed that good images were fit immediately, and that bad images never
found a good fit.

The plate scanning sequence was developed during the testing phase of the
PMM program.  The first step was to do test
exposures and adjust the neutral density filters in the illumination beam
to compensate for the background diffuse density of the plate.  This was
needed because some plates were almost clear while others were almost black.
Next, the spatial scanning pattern was done without using the imaging
cameras, but while using laser micrometers to measure the distance from
the camera to the emulsion.  This was needed because plates sagged in
the platens, the platens sagged, and because some plates had significant
ripples in the glass.  Once a 2-dimensional map of the plate surface was
measured, a low-order polynomial was fit to produce the map that was
used during the scanning to keep the CCD a constant distance from the plate.
The third step was to focus the cameras by
maximizing the sky granularity in the image.  The combination of the nominal
focus, the map of the plate surface, and the compensation for the background
density allowed the scan to commence.  Each vision system had an independent
prescan sequence, but the scanning of the plates used both cameras whenever
possible.

Each platen holds four plates, and typical, out of the Galactic plane
plates took about an hour to scan and produced about 200,000 detections.
The slowest plates were in the region of Baade's window, took about seven
hours to scan, and produced more than 7,000,000 detections.  PMM has two
platens, and barcodes were placed on all platens, plate envelopes, and
magnetic tapes to minimize bookkeeping errors.  The hardware and software
were frozen at the start of production-mode scanning (1994.7), and it
took about $2 \times 10^8$ seconds (1994.7 to 2001.2) to process essentially
all of the almost 19,000 plates involved.
About $19 \times 10^{12}$ pixels were processed (100,000 pixels per second) and
about $15 \times 10^9$ detections were logged (75 detections per second).
PMM now sits idle.

\section{Astrometric Calibration}

The first step of the calibration takes the pixel coordinates computed
in each of the 588 exposures needed to scan a plate, combines them with
the metrology of the PMM platen, and produces a catalog of positions
in PMM focal plane coordinates (integer $1 \times 10^{-8}$ meters).
It turned out that
the elaborate pre-scan efforts to determine proper focus were not
accurate enough, so a numerical refocusing algorithm was developed.
This adjusted the coordinates computed in each exposure so as to
maximize the number of overlapping frame multiple detections and
to minimize the dispersion in their mean positions.  The revision
to the focus computed by this algorithm was typically smaller than
the focal depth of the camera, but occasionally a plate was found
wherein the real time focus was significantly in error.

The second step was to compute a catalog of astrometric standards for
each plate.  To the extent enabled by the SLALIB software package
(Wallace 1994), the coordinates of stars in this catalog are
the apparent position of the star on an idealized Schmidt plate given
the circumstances of the observation.
The catalog of PMM measures was correlated with
these standards, and a cubic transformation was computed.

As is well known, the combination of the Schmidt telescope and the
photographic emulsion suffers from large astrometric errors as a
function of magnitude.  One of the problems with the astrometry
presented in USNO-A is that it used the faintest Tycho-2 stars
(H{\o}g et al. 2000) as its
astrometric calibrators.  While this was the best that could be done
at the time, all of these stars are in deep saturation on the survey
plates.  Hence, the astrometry of faint, unsaturated stars suffered
from field-dependent effects that were suspected but uncharacterized at the
time USNO-A was compiled.  A different strategy was employed for
USNO-B.  As part of the PMM program, the Northern Proper Motion
plates (Klemola et al. 1987) and the Southern Proper Motion plates
(Platais et al. 1998) were scanned, and an astrometric catalog
called YS4.0 was compiled (Monet {\em in prep.}).  The limiting
magnitude of YS4.0 is about $V=18$, and these stars are not saturated on the
Schmidt survey plates.  The advantage of this strategy is that the
astrometric calibration avoids saturated stars.  The disadvantage is
that the mean epoch of YS4.0 is about 1975, and the mean motion between
YS4.0 and the Schmidt surveys was set to zero by the least squares
solution.  Hence, the proper motions in USNO-B are relative, not
absolute, and future releases of USNO-B will attempt to remove this
bias through the introduction of a statistical model.

The first iteration of the astrometric algorithm assumed only a cubic
relationship between the plate coordinates and the astrometric standards.
Once all of the plates were processed, the astrometric residuals were
co-added as a function of position in the focal plane, and the systematic,
fixed pattern of astrometric residuals was computed.  This was done
independently for each survey, and solutions were done as a function of
magnitude and declination.  (Declination is a surrogate for zenith
distance since most plates were taken near the meridian.)  No obvious
correlations with declination were seen, but very large terms and
differences in terms were found as a function of magnitude.  All of
these maps are included in the documentation associated with the
digital version of USNO-B, but Figure 1 shows typical maps
(POSS-I 103a-O survey at internal PMM magnitudes of $m=10$ and $m=18$).
For USNO-B, separate maps were used for each survey at each integer
internal magnitude between 10 and 19.  The map for $m=10$ was used at all
internal magnitudes brighter than 10, and that for $m=19$ was used for all
magnitudes fainter than 19.

Given the maps of fixed pattern astrometric error, the astrometric
solution pipeline was rerun, the maps were included, and the cubic
mapping between plate and catalog coordinates was recomputed.  Finally,
all measures were transformed from plate coordinates to sky coordinates
to produce plate-by-plate catalogs of J2000 coordinates at the epoch
that the plate was taken.

Although a more thorough analysis of the astrometric accuracy of the
USNO-B catalog is planned, a coarse estimate can be obtained from the
dispersion of the fits for the individual observations (items {\tt Q}
and {\tt R} in Table~\ref{tbl-3}).  An analysis of the subset of
objects with all possible detections (i.e., 4 in the South and 5 in
the North; about 40\% of all objects in USNO-B) shows that the median
dispersion is about 0.12 arcsec in both $X$ and $Y$.  This is
an approximate value, and it ignores possible systematic components
arising from position on the plate, brightness of the image, or other
effects.  Users should remember that some USNO-B objects with large
astrometric uncertainty may be erroneous combinations of observations
of different objects, that the uncertainty in the predicted position
of an object increases with time since the mean epoch (often in the
1970s), and that usage of this estimated accuracy should not replace a
computation for each object using the values listed in the catalog
and the desired epoch of observation.

\section{Photometric Calibration}

The photometric calibration of the USNO-B catalog is of marginal
quality, and will be revised as the appropriate data become available.
The root of the problem is that an all-sky catalog of faint photometric
standards does not exist.  Because of this, a 2-step photometric calibration
strategy was adopted.  The first step was to calibrate the bright stars
on all the plates.  Tycho-2 stars are almost all brighter than $V=13$,
and although the images are grossly saturated, an almost
linear relationship exists between the internal magnitudes measured by
the PMM and values computed from spectral energy
distributions and the $(B,V)$ magnitudes listed in the catalog.
Details of the modeling are presented in the documentation accompanying
the digital version of the USNO-B catalog, and Henden's file
{\it tables.txt} contains the results.  Since Tycho-2 stars are measured
on all plates, the bright star fit was extrapolated and applied to all
PMM measures from that plate.

The remaining task was to compute the systematic differences between the
true calibration for faint stars and the extrapolation of the bright
star fit.  Two sources of photometry for faint stars (magnitudes from 14 to
22) were used,
the Guide Star Photometric Catalog 2 (Bucciarelli et al. 2001) and the
photometric data measured for the USNOFS CCD parallax program (Monet et
al. 1992).  Henden's tables were used to convert between the standard
system magnitudes and the photographic magnitudes, and the combination
of these two catalogs provided faint standards on 3,281 plates, or about
44\% of the total needed.  Polynomial fits (linear or cubic as
necessary) were computed, and the calibrated magnitudes for all objects
were obtained.

For the remaining 56\% of the plates without faint photometric calibrators,
the calibration was computed from the plate overlap zones.  The plates
adjacent to an uncalibrated plate were searched, and if any had a faint
calibration, then it was identified as a possible calibrator, and the
distance between plate centers was tallied.  After searching all nearby
plates, the closest one was found and if there were more than 500 objects
in common, a calibration was computed.  There were 3,517 (about 47\%)
uncalibrated plates
that were adjacent to calibrated plates, but three more steps were
needed to complete the calibration of all plates.
No analysis of all possible
overlap regions was done, and this is an area wherein improvements are
anticipated in future releases of USNO-B.

It is difficult to give a single number for the photometric accuracy
of USNO-B, but the solution combining all 632,827 calibration stars spread
over 3,281 plates has a standard deviation of 0.25 magnitudes.
This value includes whatever effects might arise from the
transformation of $(B,V,R,I)$ to $(O,E,J,F,N)$ colors, but a similar
value is found for each of the ten different surveys taken separately.
Many plates, particularly those calibrated with the photometry from the
USNOFS parallax program, show substantially smaller errors, and
solutions for plates without photometric calibrators may be worse.

Another major problem with USNO-B photometry is that a separate calibration
for the photometry of extended objects was not done.  In part, this is
due to the relatively few catalogs of galaxy photometry.  Another factor
is that the transformation from magnitudes in a standard system to those
in the photographic bandpasses is even less well determined than for
stars.  The final reason was that the photometric response of the plates
is quite non-linear, and no attempt was made to compute the transformation
between the observed values of transmitted light and the true intensity
incident on the plate.  Whereas the photometry of stars may be useful,
the USNO-B magnitudes for non-stellar objects are qualitative, at best.

\section{Star/Galaxy Separation}

Although commonly called star/galaxy separation, the value presented
in USNO-B is a measure of the similarity of the unknown object to
a stellar point spread function (PSF).  Most objects that are dissimilar
to the PSF are galaxies, but many other types of astronomical objects
and photographic artifacts can produce objects in the catalog.  Since the
measured transmission was not converted to a true intensity,  and because
the photographic process is inherently non-linear, the stellar PSF
varies with apparent brightness and position in the focal plane.

The method used to distinguish stars from nonstellar objects was an
oblique decision tree with three nodes.  Each node of the decision tree is
a plane defined by values of two image description parameters, and six
(not necessarily distinct) image parameters are needed to define this tree.
A decision tree must be trained on a set of objects of known type (star
or galaxy).  The training set in the Northern sky was chosen to be the
plates containing the Coma cluster of galaxies.  The training set was
compiled by hand by examining KPNO 4-m CCD frames of the Coma cluster
taken in good seeing by Secker (1997).  The catalog
of Coma cluster galaxies by Godwin, Metcalfe, \& Peach (1983) was also
consulted.  Because of the non-linearities in the image in transmitted
light, each interval of one internal magnitude was searched for the
set of six image description parameters that provided best star/nonstar
discrimination out of the entire set (see Table~\ref{tbl-2}).

Once the training plate of each Northern survey was calibrated, the edges
of the stellar ridge lines of the various image parameters were identified
in five annular zones centered on the plate center.  For each of the other
plates in the survey, the stellar ridge lines of the relevant parameters
were mapped (shift and stretch only) to those of the training plate zone
by zone, and the classification index was computed for each object.
The calibration of the Southern sky was done in a similar manner,
but the training set was taken to be a particularly good plate in the
equatorial overlap zone, and truth was taken to be the classification
from the Northern plates.  The largest source of classification error
is the inability to match the stellar ridge line on a given plate to
that on the training plate.

The classifier produces an integer in the range $(0-11)$ that measures
the similarity of the unknown object to a stellar PSF.  Values of
$(0-3)$ are probably non-stellar, and values of $(8-11)$ are probably
stellar.  The catalog gives the individual classifications from all
surveys other than those taken on {\em IV-N} emulsions (see
Table~\ref{tbl-3} Bytes 20--39).  Various internal tests suggest
that the classifier is correct about 85\% of the time for objects with
internal magnitudes 14--20, but a thorough discussion of the classifier,
estimates for its accuracy, and a suggested algorithm for combining the
individual estimators for a single object will be given by Canzian et al.
({\em in prep.})

\section{Catalog Compilation}

The first phase of the compilation of the final catalog is to apply
the astrometric and photometric calibrations to the raw files produced
by the PMM in real time.
The second phase is to go from a plate-based approach to files having
a constant width in South Polar Distance (SPD $= \delta + 90\arcdeg$).
It would have been nice to have preserved the 7.5-degree width used in the
USNO-A catalog, but the size of USNO-B required using a width of
0.1 degrees.  The 7,435 plate files are written into 1,800 SPD files,
and then each SPD file is sorted on right ascension.
The third phase is to examine the SPD files so that the ensemble of
individual observations is reformatted into a list of objects with
from one to five observations per object.  This task is difficult because
there are different numbers of surveys as a function of SPD, there can be
duplicate detections of the same object if it fell in a plate overlap
zone, and the epoch difference between plates is highly variable.  The
current algorithm was the following, but modifications may be needed
in future releases of USNO-B.

\begin{quote}
The first step is to find objects that do not move.  A 3.0 arcsecond aperture
is moved through the observation file for a single band of SPD, and an
object is sensed when one or more observations
are found within the aperture.  The list of observations is examined,
culled, and all single observation objects are ignored.  If the object
contains observations from one or more first epoch and one or more
second epoch surveys (see Table~\ref{tbl-1} for assignments), then
the object is tallied and the individual observations
are removed from further consideration.
\end{quote}

\begin{quote}
The second step is to look for objects with significant proper motions
that were observed at all available epochs.  The list of objects is
recomputed, omitting those that were flagged as no-motion objects, but
with the aperture of 30 arcseconds.  For each object, all combinations
of second epoch observations are fit for linear motion.  If significant, the
fit is extrapolated to the first epoch survey(s), and a search is made for
detections inside the error ellipse(s).  All possible combinations are
pursued.  If the fit includes all available surveys, has a standard
deviation less than 0.4 arcseconds in the tangent plane coordinates
$\xi$ and $\eta$, and has
a motion less than 10 (5 observations), 3 (4 observations), or 1
(3 observations) arcseconds per year, the detection is tallied and the
observations are removed from further consideration.
\end{quote}

\begin{quote}
The third step is to collect reasonable combinations of the remaining
detections into objects.  The object aperture is set to 20 arcseconds,
and all possible combinations of 5-, 4-, 3-, and 2-survey objects
are evaluated in that order.  The first collection of observations that
has a standard deviation less than 5 arcseconds in both $\xi$ and
$\eta$ is called an object, the object is tallied, and the observations
are removed from further consideration.  Only a few percent of the
objects in the USNO-B catalog come from this step, and the uncertainty
estimators in the catalog should be sufficient to identify them.
\end{quote}

\begin{quote} In all cases, the combination of individual observations
into a single object with proper motion was made without any reference
to the individual magnitudes.  Since more than 200 billion possible
combinations were considered, there may be spurious merges in the
catalog.  For many applications, a test on the individual magnitudes
can be used to identify catalog entires that may be erroneous combinations
of different objects.
\end{quote}

\begin{quote}
The final step is to remove double detections and to replace bright stars.
Double detections occur because the zones of SPD used during the compilation
are somewhat wider than 0.1 degrees to allow for edge effects.  King
and Raff (1977) give values for the relationship between image diameter
and apparent magnitude on POSS-I
plates.  Mean values were used for all surveys to remove all PMM detections
near Tycho-2 stars since the PMM's measures are usually confused by the gross
saturation, diffraction spikes, and halos.  After removing the PMM
detections, the reformatted Tycho-2 entry is inserted for completeness.
\end{quote}

Table~\ref{tbl-3} gives the data contained in USNO-B for each object.
The catalog is organized into 1,800 zones of SPD each exactly 0.1 degrees
in width, and the objects are sorted by right ascension in each zone.
In addition, accelerator files for each zone contain the first entry
and number of entries at each increment of 15 minutes of right ascension.
As was found with USNO-A, these files increase the efficiency of catalog
access.
Users who need a more detailed description are urged to read
the documentation and the source code distributed
with the catalog.  The overriding concerns were to include as many objects
as is reasonable (since the alternative is to omit the associated
detections), and to require spatially correlated detections on any
two or more surveys (even if the combinations may seem curious).
At USNOFS, the list of remaining observations has been saved, and will be
served to users upon special request.

\section{Comparison with the SDSS Early Data Release}

In an effort to substantiate the internal uncertainty estimators, the
$\sim 450$ square degrees of the Sloan Digital Sky Survey Early Data Release
(SDSS EDR; Stoughton et al. 2002) were correlated with USNO-B1.0.
The SDSS results are based on CCD data, and have substantially smaller
uncertainties than the observations from which USNO-B was compiled.
The astrometric comparison for objects determined by SDSS to be unblended
and in the magnitude range of $17 < g^* < 19$ shows a dispersion
of about 0.13 arcseconds for stars and 0.20 arcseconds for galaxies.
Unfortunately, systematic offsets as large as 0.25 arcseconds were found,
and these are taken as evidence for distortions of the USNO-B1.0
astrometric calibration.
Comparison of the stellar photometry in the magnitude range of 15 to 21
yields the following transformations and dispersions.
\begin{mathletters}
\begin{eqnarray}
 O = g^* + 0.08 + 0.452(g^*-r^*), \sigma = 0.34, \\
 E = r^* - 0.20 - 0.086(g^*-r^*), \sigma = 0.30, \\
 J = g^* + 0.06 + 0.079(g^*-r^*), \sigma = 0.33, \\
 F = r^* - 0.09 - 0.109(g^*-r^*), \sigma = 0.26, \\
 N = i^* - 0.44 - 0.164(r^*-i^*), \sigma = 0.31
\end{eqnarray}
\end{mathletters}
These relations are determined from the 34 POSS-I fields
and the 40 POSS-II fields that are covered by the SDSS EDR,
so represent only a small fraction of the USNO-B catalog.
Examination of the residuals from these relations suggests
that there are additional systematic errors that depend
on magnitude, up to 0.15 mag, and that vary from plate
to plate, up to 0.2 mag for red plates and somewhat worse
for blue plates.  Little systematic variation ($\leq$0.1 mag)
is seen with distance from the center of a plate.

Figure~\ref{fig-2} displays the fraction of SDSS objects with
matching USNO-B entries (within a radius of 1 arcsec, after applying
USNO-B proper motions to convert USNO-B positions to the epoch of the SDSS
observations) as a function of SDSS $g^*$ magnitude.
USNO-B is essentially 100\% complete for stars determined to be unblended
with other objects by SDSS, but the completeness drops to about 97\% when
all SDSS objects are considered.  This is yet more evidence that CCD data
are far superior for image deconvolution than those obtained from
photographic plates.  USNO-B is roughly 92\% complete for unblended SDSS 
galaxies with $g^* < 19.5$.
Table~\ref{tbl-4} shows the correlation between
USNO-B and SDSS classifications as a function of SDSS $g^*$ magnitude.
These accuracies are nominal since accuracy variations exist from plate to
plate and magnitude interval to magnitude interval on a given plate.

\section{Caveats and Conclusions}

USNO-B is a work in progress, and the version 1.0 public release is a
compromise between the needs of the community and the development and
testing of the catalog compilation software.  There are systematic errors and
other problems, and
future revisions will attempt to address these.  Due to its size,
human verification of every entry cannot be done.  Many of the
verification algorithms are statistical in nature, and this leaves open
the possibility for small and large errors in specific areas.  The
following, incomplete list of caveats should serve as a warning to the
users of the catalog of the types of difficulties encountered, and the
need for skepticism in believing the attributes of peculiar entries.

\begin{quote}
The photometric calibration is preliminary.  Only 44\% of the fields have
faint calibrators somewhere on the plate, and in many cases there are
only a very few of these.  There is no separate calibration of the magnitudes
of non-stellar objects, so these should be treated as qualitative measures.
The catalog presents magnitudes in the photographic systems $(O,E,J,F,N)$
and not on standard systems to minimize uncertainties and the chances for
systematic errors.  A vignetting function for the Schmidt telescopes has
not been studied beyond the preliminary tests that showed that it was
probably unimportant.  Photometry is probably the weakest aspect of
USNO-B, and will be the subject of the most work in the future.
\end{quote}

\begin{quote}
Please do not ignore the proper motions.  The catalog lists the positions
for objects in J2000 at the epoch 2000.0, and in many cases this epoch is
more than 25 years away from the mean epoch of the observations.
Errors in the proper motion, or errors in assigning particular observations
to the same object that give rise to spurious proper motions produce
catalog positions for patches of empty sky.  If you wish to ignore the true
proper motion, you should use the catalog motions to compute the position at
the mean epoch and use this position instead of the 2000.0 position.  This is
particularly important for objects with large proper motions ($\mu \ge 1.0$
aresec/year).  The epoch difference of the plate material in a single
field can often be as large as 50 years.  Allowing the software to use
a large search radius greatly increases the probability for spurious
detections, but limiting the search radius to a small value will remove all
real high proper motion objects from the catalog.  Users should understand that
the reality of grouping observations into a single object with
high proper motion is a strong function of the number and epoch of the
individual observations.  Extreme caution is suggested for all such objects,
particularly those that are missing one or more opportunities for detection
due to faint magnitude or extreme color.
\end{quote}

\begin{quote}
USNO-B presents relative, not absolute, proper motions.  The calibration
process described above sets the mean motion of all objects correlated
with entries in the YS4.0 catalog to be zero.  Since least squares was
used, the solution is dominated by large numbers of faint stars, and it
is believed that the mean correlated star is between yellow magnitudes
of 17 and 18. Whereas this motion may be small in absolute terms, it
may be significant in various statistical studies.  Future releases of
USNO-B will attempt to minimize or remove this bias.
\end{quote}

\begin{quote}
USNO-B covers the entire sky, and users should understand that there are
many areas of the sky that do not resemble sparse, high Galactic latitude
fields where objects are isolated and the sky is well defined.  In dense
fields, the sky is only poorly determined, and PMM often counted more than
5 million distinct objects on a Schmidt plate.  Users should understand that
photographic photometry and astrometry are not well defined in such
regions, nor are they well defined in regions of nebulosity, well-resolved
objects like globular clusters and bright galaxies, and around bright stars.
The ``any 2'' rule for merging observations from individual surveys into
objects arose from the desire to retain faint blue or faint red objects
that were omitted from USNO-A, but the penalty is to increase the probability
of false objects in peculiar regions of sky.  USNO-B data should not be used
for the study of such regions:  67 arcsec/mm is a very coarse scale, the
dynamic range of plates is limited, and PSF reconstruction with multi-PSF
image deconvolution was not done.  Again, this catalog is based on
photographic data, and these are of lower quality than those that are familiar
to younger astronomers.
\end{quote}

\begin{quote}
Please do not use USNO-B as a source of data for bright stars.  For the
sake of completeness and the minimization of confusion, holes were cut
around Tycho-2 stars and the Tycho-2 data were copied into the catalog.
The original Tycho-2 data should be used for all critical applications.
Stars within a few magnitudes of the Tycho-2 limiting magnitude are
saturated on the Schmidt plates, and astrometric and photometric values
are of lower accuracy.
\end{quote}

USNO-B is one of the few attempts to digitize, process, and analyze the
entire sky as seen in different optical colors and at different epochs, and
it contains successes and failures.
It is an all-sky catalog of positions, proper motions, magnitudes,
and classifications for more than a billion objects, and its accuracy
is about 0.2 arcseconds for astrometry, 0.3 magnitudes for photometry,
and 85\% for object classification.
As demonstrated
by the comparison to the SDSS EDR and other zone catalogs, systematic
astrometric and photometric errors exist in this version, and there may be
large regions of
sky wherein the automated verification algorithms failed without signaling
an error.  The release of USNO-B1.0 is a compromise between the desire
for public access and the process of catalog compilation and verification.
The next step in the USNO-B process will be the database join with
the 2-Micron All Sky Survey (Skrutskie et al. 1997;
{\em http://www.ipac.caltech.edu/2mass}), and this will entail a
top-to-bottom verification and recalibration of the USNO-B algorithms.  The
scheduled release date for USNO-B1.1 is September 2003.
As was USNO-A, USNO-B is just a milestone in the processing of the PMM's
archive.  Future work will be directed toward finding a way
to calibrate the data from all of the plates that PMM has scanned, and
to identify a method to make these available to the community.  Manpower
and costs are finite, but the quality and importance of the historic
photographic data justify these efforts.

\section{Acknowledgments}

The first, and probably the most important acknowledgment goes
to the many individuals (observers, technicians, assistants, graduate
students, etc.), mostly anonymous except in biodegrading paper logs
and documents, whose dedication and expertise have produced the
stunning collection of plates that PMM had the privilege of scanning.
The record of the sky is irreplaceable, and that it is of such high
quality serves as an inspiration to those who follow in their footsteps.
Most of the errors and inaccuracies in USNO-B belong to PMM, and not
to the original archive.

Although technically in the public domain and hence requiring no
acknowledgment, it is appropriate to repeat these sentences from
Minkowski and Abell (1963). ``The Sky Survey was made financially
possible by grants from the National Geographic Society.  The society
provided the photographic materials and special equipment required,
the salaries of the personnel employed full or part time on the
survey, and the production of the two sets of contact positives on glass
of each survey photograph.  The observing time with the 48-inch Schmidt
telescope required to obtain the Sky Survey photographs was made available
by the Palomar Observatory of the California institute of Technology.''

This work is based partly on photographic plates obtained at the Palomar
Observatory 48-inch Oschin Telescope for the Second Palomar
Observatory Sky Survey which was funded by the Eastman Kodak
Company, the National Geographic Society, the Samuel Oschin
Foundation, the Alfred Sloan Foundation, the National Science
Foundation grants AST84-08225, AST87-19465, AST90-23115 and
AST93-18984,  and the National Aeronautics and Space Administration
grants NGL 05002140 and NAGW 1710.

Some of the measures used in the USNO-B catalog are based
on photographic data obtained using
the UK Schmidt Telescope. The UK Schmidt Telescope was operated by the
Royal Observatory Edinburgh, with funding from the UK Science and
Engineering Research Council, until June 1988, and thereafter by the
Anglo-Australian Observatory. Original materials are copyrighted by
the Royal Observatory Edinburgh and the Anglo-Australian Observatory, and the
plates were scanned with their permission.

Thanks are expressed to the European Southern Observatory (ESO) for permission
to scan the ESO-R survey glass plate copies housed at the National
Optical Astronomy Observatory.  ESO retains copyright and other intellectual
property rights to these plates.

Special thanks are extended to those at California Institute of
Technology, Royal Observatory Edinburgh, and the Anglo-Australian
observatory for the loan of original plate material, and to
the National Optical Astronomy Observatory for loaning their collection
of glass copies.

The continuing support of the U. S. Air Force, particularly HQ AFSPC/DOY
(and predecessors) and Dr. Joseph Liu, is greatly appreciated.  It is
rare to see such a successful combination of Mission, Operational, and
scientific goals.

BC is grateful to Jeff Secker for providing invaluable high resolution
digital imaging of the Coma cluster with which the training set for the
image classifier was constructed.

Funding for the creation and distribution of the SDSS Archive has been provided
by the Alfred P. Sloan Foundation, the Participating Institutions, the National
Aeronautics and Space Administration, the National Science Foundation, the U.S.
Department of Energy, the Japanese Monbukagakusho, and the Max Planck Society.
The SDSS Web site is {\em http://www.sdss.org/}.

The SDSS is managed by the Astrophysical Research Consortium (ARC) for the
Participating Institutions. The Participating Institutions are The University
of Chicago, Fermilab, the Institute for Advanced Study, the Japan Participation
Group, The Johns Hopkins University, Los Alamos National Laboratory,
the Max-Planck-Institute for Astronomy (MPIA), the Max-Planck-Institute for
Astrophysics (MPA), New Mexico State University, University of Pittsburgh,
Princeton University, the United States Naval Observatory, and the University
of Washington.

\begin{deluxetable}{lcccccc}
\tabletypesize{\scriptsize}
\tablecaption{Photographic data. \label{tbl-1}}
\tablewidth{0pt}
\tablehead{
\colhead{Survey} &
\colhead{Emulsion} &
\colhead{Wavelength (nm)} &
\colhead{Declination\tablenotemark{a}} &
\colhead{Fields} &
\colhead{Epoch\tablenotemark{b}} &
\colhead{Archive\tablenotemark{c}}
}
\startdata
POSS-I & 103a-O & 350-500 & $-$30 - $+$90 & 936 & 1949-1965 ($1^{st}$) & O \\
POSS-I & 103a-E & 620-670 & $-$30 - $+$90 & 936 & 1949-1965 ($1^{st}$) & O \\
POSS-II & IIIa-J & 385-540 & 0 - $+$87.5 & 897 & 1985-2000 ($2^{nd}$) & O \\
POSS-II & IIIa-F & 610-690 & 0 - $+$87.5 & 897 & 1985-1999 ($2^{nd}$) & O \\
POSS-II & IV-N & 730-900 & $+$5 - $+$87.5 & 800 & 1989-2000 & O \\
SERC-J & IIIa-J & 395-540 & $-$90 - $-$20 & 606 & 1978-1990 ($2^{nd}$) & G \\
SERC-EJ & IIIa-J & 395-540 & $-$15 - $-$5 & 216 & 1984-1998 ($2^{nd}$) & G \\
ESO-R & IIIa-F & 630-690 & $-$90 - $-$35 & 408 & 1974-1987 ($1^{st}$) & G \\
AAO-R & IIIa-F & 590-690 & $-$90 - $-$20 & 606 & 1985-1998 ($2^{nd}$) & O \\
SERC-ER & IIIa-F & 590-690 & $-$15 - $-$5 & 216 & 1979-1994 ($2^{nd}$) & G \\
SERC-I & IV-N & 715-900 & $-$90 - 0 & 892 & 1978-2002 & O \\
SERC-I\tablenotemark{d} & IV-N & 715-900 & $+$5 - $+$20 & 25 & 1981-2002 & O \\
 \enddata
\tablecomments{References for the individual surveys can be found in
Morgan et al. (1992), and the wavelength data presented above were copied
from this source.}
\tablenotetext{a}{Range of field centers in nominal B1950 used in the
compilation of USNO-B.  In many cases, the survey covers a larger area.}
\tablenotetext{b}{The assignment of $1^{st}$ or $2^{nd}$ epoch is
used in the compilation of the catalog.  In this notation, there is
no $1^{st}$ epoch blue survey south of $\delta = -33 \arcdeg$.}
\tablenotetext{c}{O is the original plate, G is the glass copy.  In isolated
cases, a glass copy was scanned when the original plate was found to be
broken, missing, or otherwise unacceptable.  For surveys marked with ``G'',
the glass copies were borrowed from the National
Optical Astronomy Observatory.}
\tablenotetext{d}{Extension of SERC-I survey to fill holes in POSS-II IV-N
survey.}
\end{deluxetable}
\begin{deluxetable}{lll}
\tabletypesize{\scriptsize}
\tablecaption{PMM Image Parameters.  \label{tbl-2}}
\tablewidth{0pt}
\tablehead{
\colhead{Name} &
\colhead{Description} &
\colhead{Range of Values}
}
\startdata
NFit & Total number of pixels in fit & 0 - 9999 \\
NSat & Number of saturated pixels in fit & 0 - 9999 \\
CCDX & Integer X (column) pixel & 0 - 9999 \\
CCDY & Integer Y (row) pixel & 0 - 9999 \\
Frame & Frame count into scan & 0 - 999 \\
$A$ & Amplitude of fit & $-$99.9 - 899.9 \\
$B$ & Background of fit & $-$99.9 - 899.9 \\
$x_0$ & $X$ position on plate from fit & 0.00 - 999999.99 \\
$y_0$ & $Y$ position on plate from fit & 0.00 - 999999.99 \\
$r_0^2$ & Saturation radius from fit & 0.0 - 999.9 \\
$\alpha$ & Wing scale length from fit & 0.0 - 999.9 \\
$\sigma$ & Standard deviation of fit & 0.0 - 999.9 \\
Mag & LOG10(Sum of flux in pixels in fit) & 0.00 - 99.99 \\
M9 & LOG10(Sum of flux in central 3x3 pixels) & 0.00 - 99.99 \\
M00 & Image moment LOG10($\Sigma wx^0y^0$) & 0.00 - 9.99 \\
M10 & Image moment LOG10($\Sigma wx^1y^0$) & 0.00 - 9.99 \\
M01 & Image moment LOG10($\Sigma wx^0y^1$) & 0.00 - 9.99 \\
M20 & Image moment LOG10($\Sigma wx^2y^0$) & 0.00 - 9.99 \\
M11 & Image moment LOG10($\Sigma wx^1y^1$) & 0.00 - 9.99 \\
M02 & Image moment LOG10($\Sigma wx^0y^2$) & 0.00 - 9.99 \\
MR$+$1 & Image moment LOG10($\Sigma wr^{+1}$) & 0.00 - 9.99 \\
MR$-$1 & Image moment LOG10($\Sigma wr^{-1}$) & 0.00 - 9.99 \\
R$+$$+$ & Image radius in $(+X,+Y)$ direction & 0 - 99 \\
R$+$$-$ & Imgae radius in $(+X,-Y)$ direction & 0 - 99 \\
R$-$$+$ & Image radius in $(-X,+Y)$ direction & 0 - 99 \\
R$-$$-$ & Image radius in $(-X,-Y)$ direction & 0 - 99 \\
Grad & Maximum image gradient (DN/pixel) in image & 0 - 999 \\
Lump & Dispersion of DN values in image core & 0 - 999 \\
Blend & Identifier for blended image & 0 - 9999 \\
Depth & Depth of subdivided blend & 0 - 9999 \\
\enddata
\tablecomments{DN is the raw image data number (0 - 255) produced by
the camera, and $w$ is (DN-Background).
The signs of the arguments of the logarithms were saved separately.
All quantities were computed for every
image, and were packed into 52 bytes.  The range of values shows the
quantization associated with the packing, and not the precision of the
calculation.  For further details, please consult the
source code distributed with the catalog.
}
\end{deluxetable}
\begin{deluxetable}{rrrlcc}
\tabletypesize{\scriptsize}
\tablecaption{Data for Each Entry in USNO-B. \label{tbl-3}}
\tablewidth{0pt}
\tablehead{
\colhead{Bytes} &
\colhead{Packing\tablenotemark{a}} &
\colhead{Field} &
\colhead{Description} &
\colhead{Quanta} &
\colhead{Decimal Range}
}
\startdata
0 - 3 & {\tt aaaaaaaaa} & & J2000 epoch 2000.0 RA & 0.01 arcsec &
0.00 - 1,295,999.99 arcsec \\
4 - 7 &  {\tt ssssssss} & & J2000 epoch 2000.0 SPD & 0.01 arcsec &
0.00 - 648,000.00 arcsec \\
8 - 11 & {\tt iPSSSSAAAA} & {\tt AAAA} & $\mu_{RA}$\tablenotemark{b} &
0.002 arcsec/year & $-$10.000 - $+$9.998 arcsec/year \\
  &            & {\tt SSSS} & $\mu_{SPD}$\tablenotemark{b} &
0.002 arcsec/year & $-$10.000 - $+$9.998 arcsec/year \\
  &            &  {\tt P}   & total $\mu$ probability & 0.1 & 0.0 - 0.9 \\
  &            &  {\tt i}   &
\multicolumn{3}{l}{motion catalog flag: 0=no, 1=yes\tablenotemark{c}} \\
12 - 15 & {\tt jMRQyyyxxx} & {\tt xxx}  &
$\sigma_{\mu_{RA}}$\tablenotemark{b} & 0.001 arcsec/year &
0.000 - 0.999 arcsec/year \\
  &            & {\tt yyy}  &
$\sigma_{\mu_{SPD}}$\tablenotemark{b} & 0.001 arcsec/year &
0.000 - 0.999 arcsec/year \\
  &            &  {\tt Q}   &
$\sigma_{RA fit}$\tablenotemark{b} & 0.1 arcsec & 0.0 - 0.9 arcsec \\
  &            &  {\tt R}   &
$\sigma_{SPD fit}$\tablenotemark{b} & 0.1 arcsec & 0.0 - 0.9 arcsec \\
  &            &  {\tt M}   & Number of detections & & 2 - 5\tablenotemark{d} \\
  &            &  {\tt j}   &
\multicolumn{3}{l}{diffraction spike flag: 0=no, 1=yes\tablenotemark{e}} \\
16 - 19 & {\tt keeevvvuuu} & {\tt uuu} &
$\sigma_{RA}$\tablenotemark{b} & 0.001 arcsec &
0.000 - 0.999 arcsec \\
   &            & {\tt vvv} &
$\sigma_{SPD}$\tablenotemark{b} & 0.001 arcsec & 0.000 - 0.999 arcsec \\
   &            & {\tt eee} & mean epoch $-$ 1950.0 & 0.1 year &
1950.0 - 2049.9 \\
   &            &  {\tt k}  &
\multicolumn{3}{l}{YS4.0 correlation flag: 0=no, 1=yes\tablenotemark{f}} \\
20 - 23 & {\tt GGSFFFmmmm} & {\tt mmmm} & $1^{st}$ blue magnitude & 0.01 mag &
0.00 - 99.99 mag \\
        &            &  {\tt FFF} & $1^{st}$ blue field & &
1 - 937\tablenotemark{g} \\
        &            &   {\tt S}  & $1^{st}$ blue survey & &
0 - 9\tablenotemark{h} \\
        &            &   {\tt GG} & $1^{st}$ blue star/galaxy & &
0 - 11\tablenotemark{i} \\
24 - 27 & {\tt GGSFFFmmmm} & \multicolumn{4}{l}{same as above except for
$1^{st}$ red survey} \\
28 - 31 & {\tt GGSFFFmmmm} & \multicolumn{4}{l}{same as above except for
$2^{nd}$ blue survey} \\
32 - 35 & {\tt GGSFFFmmmm} & \multicolumn{4}{l}{same as above except for
$2^{nd}$ red survey} \\
36 - 39 & {\tt GGSFFFmmmm} & \multicolumn{4}{l}{same as above except for
$N$ survey} \\
40 - 43 & {\tt CrrrrRRRR} & {\tt RRRR} &
$1^{st}$ blue $\xi$ residual\tablenotemark{b} & 0.01 arcsec &
$-$50.00 - $+$49.99 arcsec \\
        &          & {\tt rrrr} &
$1^{st}$ blue $\eta$ residual\tablenotemark{b} & 0.01 arcsec &
$-$50.00 - $+$49.99 arcsec \\
 & & C & \multicolumn{2}{l}{source of photometric calibration} &
0  - 9\tablenotemark{j} \\
44 - 47 & {\tt CrrrrRRRR} & \multicolumn{4}{l}{same as above except for
$1^{st}$ red survey} \\
48 - 51 & {\tt CrrrrRRRR} & \multicolumn{4}{l}{same as above except for
$2^{nd}$ blue survey} \\
52 - 55 & {\tt CrrrrRRRR} & \multicolumn{4}{l}{same as above except for
$2^{nd}$ red survey} \\
56 - 59 & {\tt CrrrrRRRR} & \multicolumn{4}{l}{same as above except for
$N$ survey} \\
60 - 63 & {\tt iiiiiii} & & \multicolumn{3}{l}{$1^{st}$ blue lookback index
into PMM scan file} \\
64 - 67 & {\tt iiiiiii} & \multicolumn{4}{l}{same as above except for
$1^{st}$ red survey} \\
68 - 71 & {\tt iiiiiii} & \multicolumn{4}{l}{same as above except for
$2^{nd}$ blue survey} \\
72 - 75 & {\tt iiiiiii} & \multicolumn{4}{l}{same as above except for
$2^{nd}$ red survey} \\
76 - 79 & {\tt iiiiiii} & \multicolumn{4}{l}{same as above except for
$N$ survey} \\
 & & & & & \\
 & & & & & \\
 & & & & & \\
 & & & & & \\
 & & & & & \\
 & & & & & \\
 & & & & & \\
 & & & & & \\
 & & & & & \\
 & & & & & \\
 & & & & & \\
 & & & & & \\
 & & & & & \\
 & & & & & \\
\enddata

\tablecomments{To preserve a fixed length record for an arbitrary set
of detections, the fields associated with a missing survey in the three
survey dependent segments (bytes 20 - 39, 40 - 59, 60 - 79) are set
to zero.  The number of non-zero entries in each segment will agree with
the number of detections {\tt M} contained in bytes 12 - 15.  Tycho-2
star entries have the same length and a similar (but not identical) format.}
\tablenotetext{a}{Fields seen when the packed integer is printed as a base-10
integer.  Combinations of integer division and subtraction are used to
extract a particular field.}
\tablenotetext{b}{The calculation for position and motion is done in
the tangent plane using the standard coordinates $\xi$ and $\eta$ with
the mean position taken as the tangent point.  This is an approximation,
and will be incorrect for the most demanding of applications.}
\tablenotetext{c}{Objects with large proper motions were cross-correlated
with the LHS (Luyten 1979a), the NLTT (Luyten 1979b), and the Lowell
Proper Motion Survey (Giclas et al. 1971, Giclas et al. 1978).
This flag is set if the
USNO-B detection has a correlation in any of these catalog, but the data
presented are from the PMM and not from the proper motion catalog.}
\tablenotetext{d}{$M=0$ denotes Tycho-2 stars, $M=1$ objects are
omitted from USNO-B but saved in catalog of rejects.}
\tablenotetext{e}{Diffraction spikes from bright stars cause spurious
PMM detections.  Occasionally, these will correlate to produce a
spurious object.  The distance to and bearing of the nearest
Tycho-2 star was computed, and the flag is set if the distance is
within 5 times the radius computed by King and Raff (1977) and the
bearing is within 5 degrees of the cardinal points (0, 90, 180, and 270
degrees).}
\tablenotetext{f}{This object was correlated with an object in the
YS4.0 catalog (Monet {\em in prep.}), and therefore contributes
in the solution for the mean proper motion of the field.}
\tablenotetext{g}{Field number in original survey as shown in
Table~\ref{tbl-1}.}
\tablenotetext{h}{Surveys encoded as 0=POSS-I $O$, 1=POSS-I $E$,
2=POSS-II $J$, 3=POSS-II $F$, 4=SERC-J or SERC-EJ, 5=ESO-R or SERC-ER,
6=AAO-R, 7=POSS-II $N$, 8=SERC-I, 9=POSS-II $N$ field number but plate
taken as part of SERC-I.}
\tablenotetext{i}{The star/galaxy estimator is a measure of the similarity
of the image to a stellar point spread function. 0 means quite dissimilar and
11 means quite similar.}
\tablenotetext{j} {Photometric calibration source: 0=bright photometric
standards on plate, 1=faint photometric standards 0 plates away (i.e.,
on this plate), 2=faint photometric
standards 1 plate away (i.e., on overlap plate), 3=faint photometric standards
2 plates away (i.e., on overlap of overlap plate), etc.}
\end{deluxetable}

\begin{deluxetable}{llllllllll}
\tabletypesize{\scriptsize}
\tablecaption{Agreement Between USNO-B and SDSS Image Classification. \label{tbl-4}}
\tablewidth{0pt}
\tablehead{
\colhead{Survey} &
\multicolumn{9}{c}{$g$ magnitude}\\
 & \colhead{13} & \colhead{14} & \colhead{15} & \colhead{16} &
 \colhead{17} & \colhead{18} & \colhead{19} & \colhead{20} &
 \colhead{$\geq 21$}\\
}
\startdata
\multicolumn{10}{c} {188,515 Stars} \\
POSS-I $O$ & 0.50 & 0.81 & 0.78 & 0.93 & 0.96 & 0.88 & 0.74 & 0.49 & 0.31 \\
POSS-I $E$ & 0.67 & 0.85 & 0.84 & 0.97 & 0.96 & 0.95 & 0.92 & 0.80 & 0.53 \\
POSS-II $J$ & 0.50 & 0.80 & 0.84 & 0.89 & 0.92 & 0.92 & 0.79 & 0.55 & 0.37 \\
POSS-II $F$ & 1.00 & 0.97 & 0.93 & 0.94 & 0.95 & 0.93 & 0.78 & 0.59 & 0.52 \\
\tableline
\multicolumn{10}{c} {172,020 Galaxies} \\
POSS-I $O$ & 0.24 & 0.36 & 0.88 & 0.92 & 0.89 & 0.95 & 0.76 & 0.71 & 0.75 \\
POSS-I $E$ & 0.33 & 0.33 & 0.84 & 0.82 & 0.82 & 0.87 & 0.85 & 0.82 & 0.86 \\
POSS-II $J$ & 0.17 & 0.21 & 0.89 & 0.96 & 0.91 & 0.83 & 0.83 & 0.85 & 0.84 \\
POSS-II $F$ & 0.21 & 0.25 & 0.84 & 0.94 & 0.91 & 0.83 & 0.91 & 0.90 & 0.88 \\
\enddata
\end{deluxetable}

\begin{figure}
\plottwo{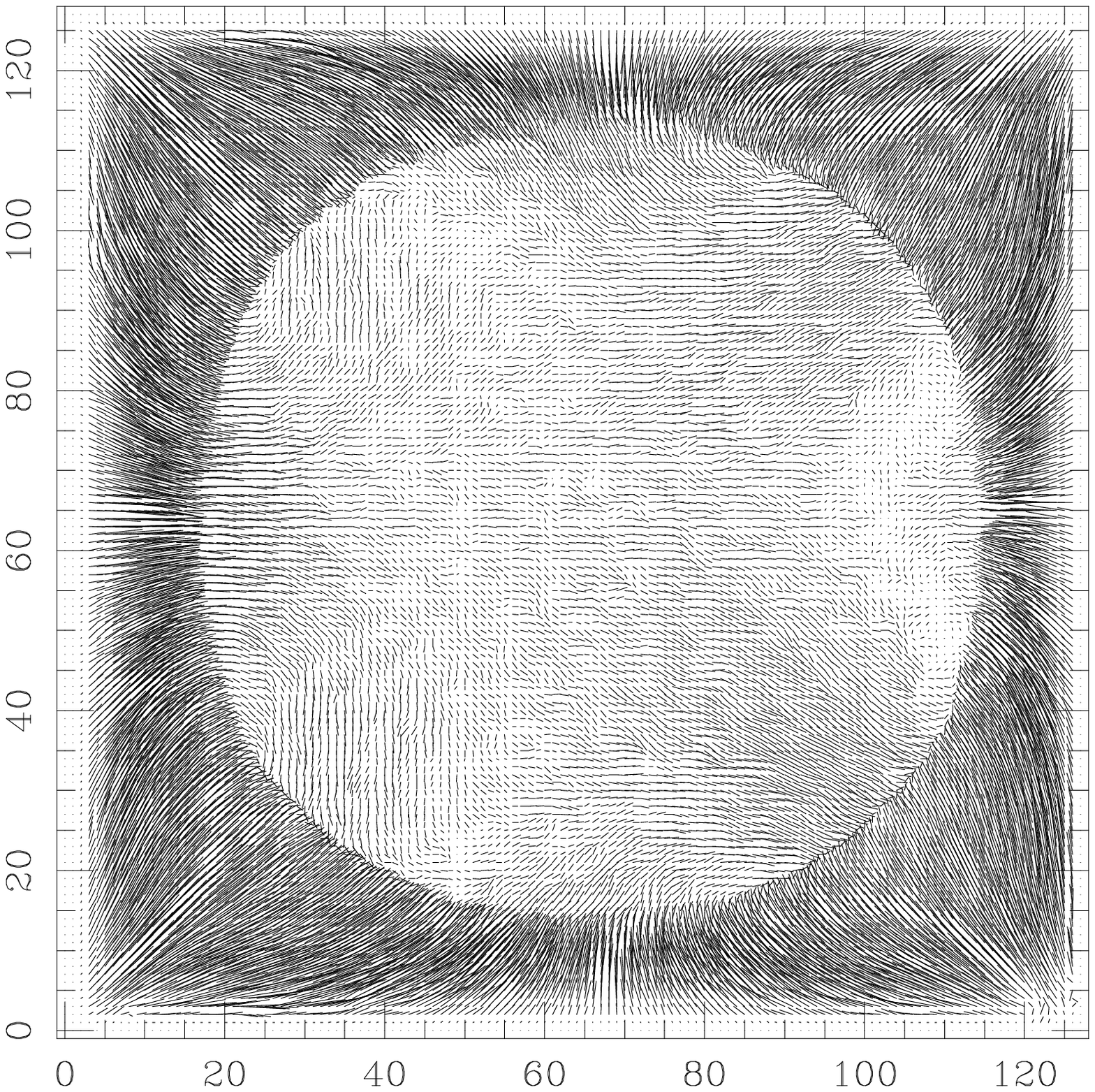}{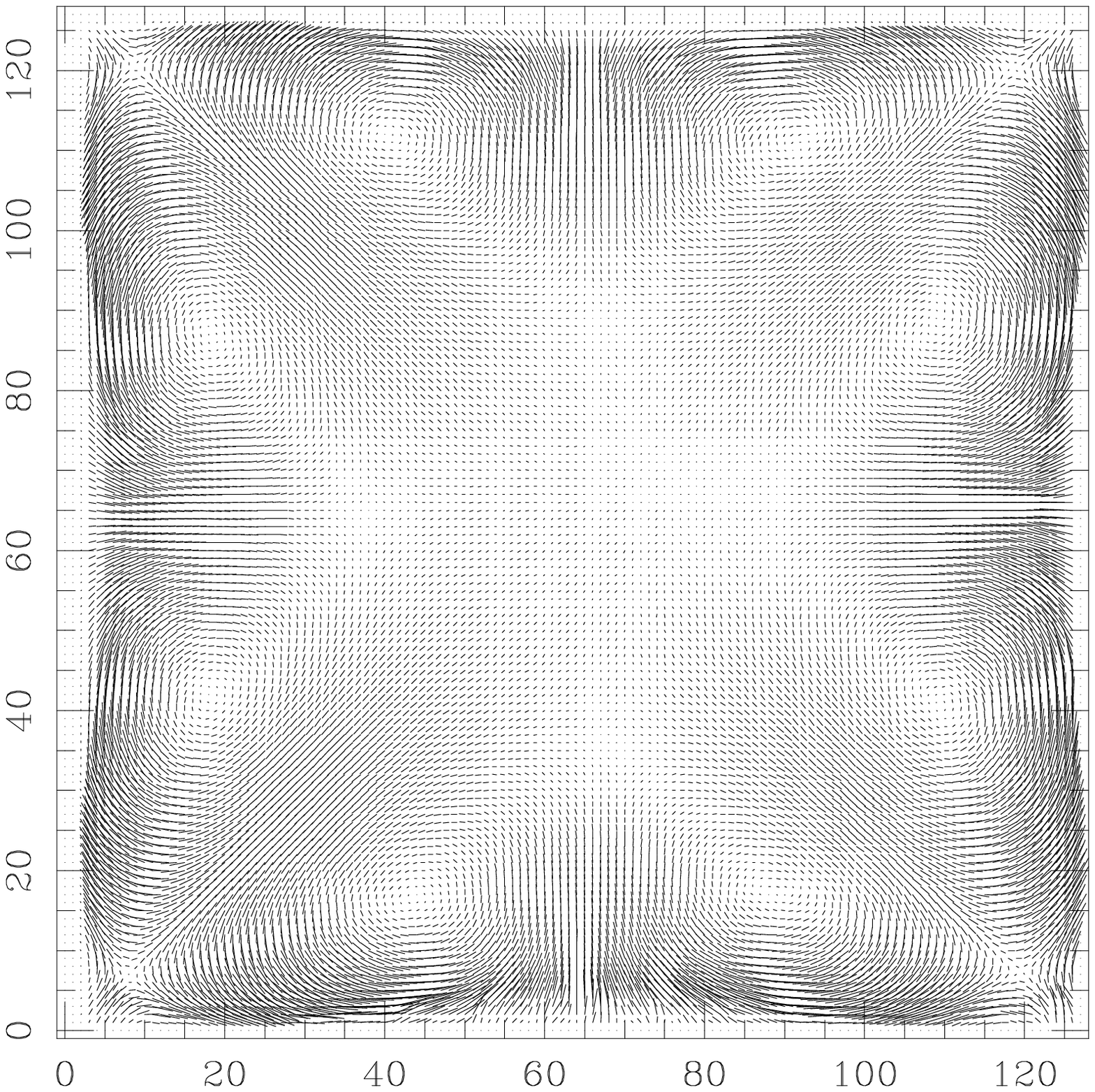}
\caption{
Maps of fixed pattern astrometric residuals for the mean POSS-I $O$ survey
plate at $10^{th}$ (left) and $18^{th}$ (right) magnitude.  The diagrams
are labeled in bins of 5 millimeter width, and the mean residuals are scaled
such that 0.20 arcseconds is the spacing between bins.
\label{fig1}}
\end{figure}

\begin{figure}
\plotone{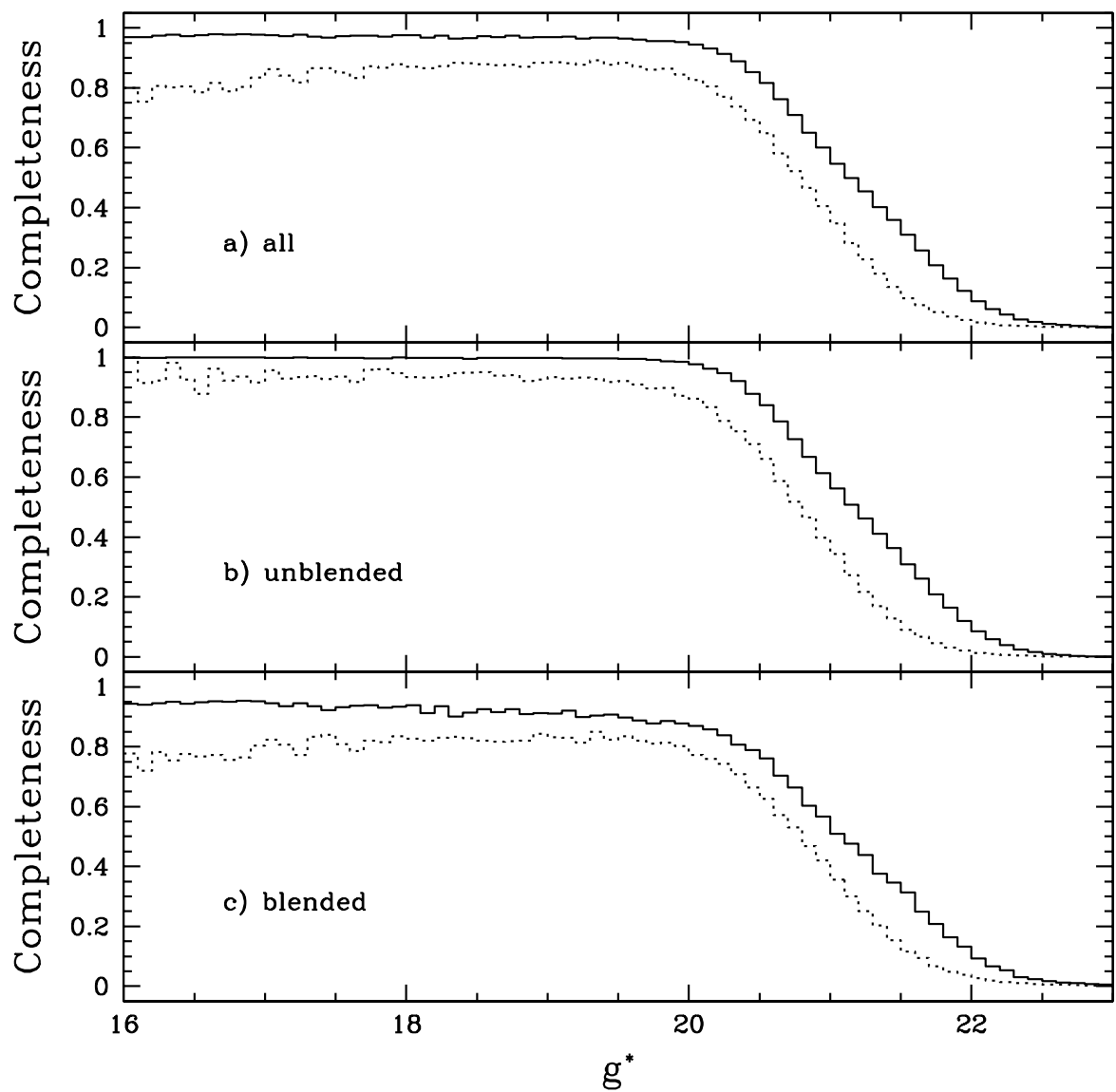}
\caption{Fraction of SDSS objects detected by USNO-B as a function of
SDSS $g^*$ magnitude.  The solid and dashed lines are for objects classified
by SDSS as stars and galaxies, respectively.  Panel a shows all objects, panel
b shows objects determined to be unblended by SDSS, and panel c
shows objects determined to be blended with other objects by SDSS.
\label{fig-2}}
\end{figure}

\end{document}